\documentclass[twocolumn,letterpaper,amsmath,amssymb]{revtex4}
\usepackage{graphicx}
\usepackage{dcolumn}
\usepackage{bm}
\usepackage{color}

\newcommand{\beq}[1]{  \begin{equation} \label{#1} }
\newcommand{\beqa}[1]{\begin{eqnarray} \label{#1} }

\newcommand{\eeq}{\end{equation}}
\newcommand{\eeqa}{\end{eqnarray}}

\begin{document}

\newcommand{\rf}[1]{(\ref{#1})}

\newcommand{\bfomega}{ \mbox{\boldmath{$\omega$}}}

\title{
Dynamic effective mass of granular media}

\author{ Chaur-Jian Hsu$^1$, David L. Johnson$^1$, Rohit
A. Ingale$^2$, John J. Valenza$^1$, Nicolas Gland$^2$, and Hern\'an
A. Makse$^2$}

\affiliation{
$^1$Schlumberger-Doll Research, One Hampshire Street,
Cambridge, MA 02139\\
$^2$Levich Institute and Physics Department, City College of New York, New
York, NY 10031}
\date{\today}

\begin{abstract}
  We develop the concept of frequency dependent effective mass,
  $\tilde{M}(\omega)$, of jammed granular materials which occupy a
  rigid cavity to a filling fraction of 48\%, the remaining volume
  being air of normal room condition or controlled humidity.  The
  dominant features of $\tilde{M}(\omega)$ provide signatures of the
  dissipation of acoustic modes, elasticity and aging effects in the
  granular medium. We perform humidity controlled experiments and
  interpret the data in terms of a continuum model and a ``trap''
  model of thermally activated capillary bridges at the contact
  points.  The results suggest that attenuation in the granular
  materials studied here can be influenced significantly by the
  kinetics of capillary condensation between the asperities at the
  contacts.
  \linebreak\linebreak{PACS: 45.70.-n, 46.40.-f, 81.05.Rm}
\end{abstract}



\maketitle



A distinct feature of jammed or loosely packed granular materials
made of a variety of different materials such as sand, steel,
polymer or glass is the ability to dissipate acoustic energy
through the network of interparticle contacts or viscous
dissipation through the surrounding medium. Indeed, loose grains
damp acoustic modes very efficiently
\cite{cremer,duran,liu,bourinet,bourinet2} and they are routinely
used as an effective method to optimize the damping of unwanted
structure-borne acoustic signals \cite{cremer}. Despite its
fundamental importance and practical applications, the microscopic
origins of the mechanisms of dissipation in jammed granular
materials are still unknown.

In this Letter, we pursue the concept of a frequency dependent
effective mass, $\tilde{M}(\omega)$, of a loose granular aggregate
contained within a rigid cavity \cite{bourinet2}. The effective
mass $\tilde{M}(\omega)$ is complex valued; its real part reflects
the inertial and elastic properties while its imaginary part
reflects the dissipative properties of the granular medium. We
demonstrate how the features of $\tilde{M}(\omega)$ allow the
study of some of the mechanisms of damping of acoustic modes,
aging and elasticity in granular matter.


Generally speaking, the real part of $\tilde{M}(\omega)$ exhibits
a sharp resonance which we interpret in terms of an effective
sound speed. The imaginary part of $\tilde{M}(\omega)$ shows a
broad resonance peak which quantifies the attenuation of acoustic
waves in the system. We observe significant changes in the
stiffness and attenuation as a function of humidity. By monitoring
the effective mass in time, we find a logarithmic aging effect in
the resonance frequency as well as an increase of the damping upon
humidification.
These effects can be modeled as capillary condensation occurring
between the asperities at the contact surfaces between the grains
during humidity-dry cycles. We interpret this phenomenon in the
context of a ``trap model'' of thermally activated liquid bridges.
Our results suggest that, in the granular materials investigated
in the present study, dissipation of acoustic energy is dominated
by the asperities at the interparticle contact surfaces. In
addition, humidity drastically affects the attenuation of the
material through the capillary condensation of liquid bridges.

\begin{figure}[b]
\centerline{ \hbox{ \resizebox{8cm}{!} {
\includegraphics{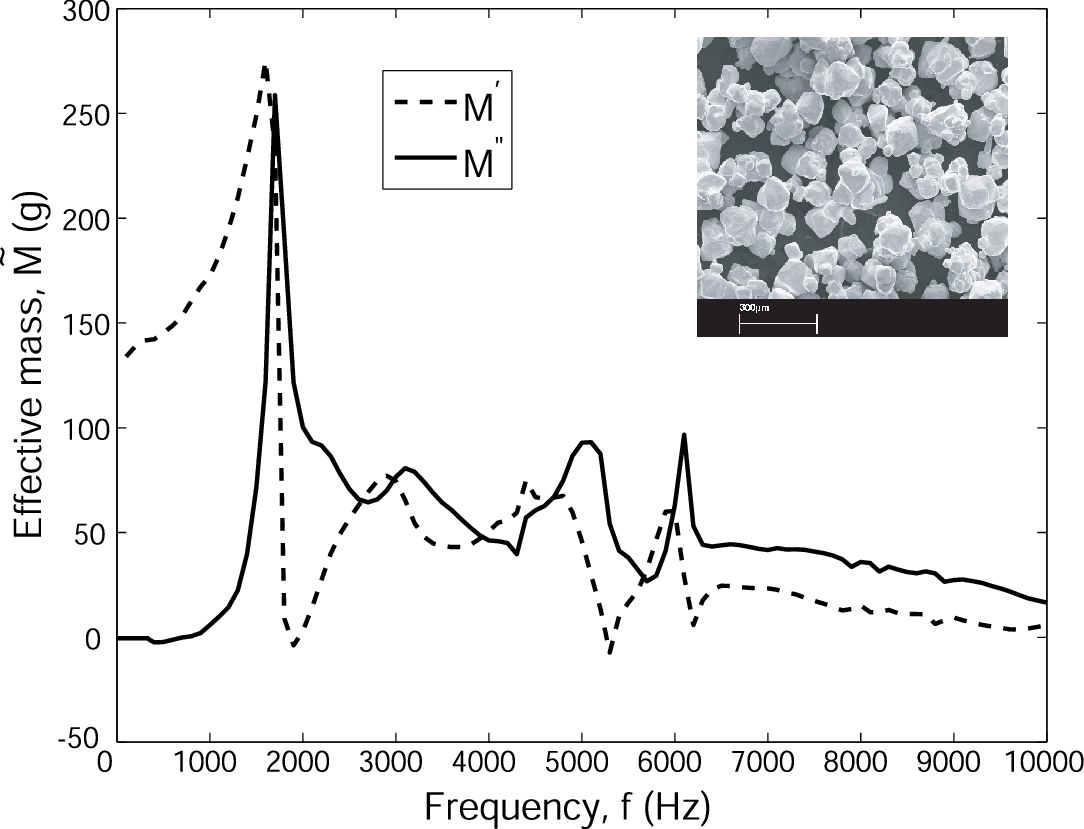}} } }
\caption {Effective mass of tungsten particles in a rigid cavity
as a function of driving frequency. Inset: Scanning Electron
Microscope image of the tungsten particles used in our
experiments. The scale at the bottom of the image corresponds to
$300 \mu$m.} \label{experiments}
\end{figure}

{\it Experiments.---} A cylindrical cavity (of diameter $D$=25.4 mm)
excavated in a rigid aluminum cup is filled with tungsten particles up
to a height of 30.7 mm.  We use tungsten particles due to their large
density, which maximizes the effects we are studying.  Each of these
particles consists of four or five equal-axis particles, of nominal
size 100 $\mu$m, fused together (see inset of
Fig. \ref{experiments}). Such irregular arrangement produces low
volume fractions (in this case $48\%$, in comparison with random close
packing of spherical particles, $64\%$) and allows to study the loose
packings of interest in the present work.  To investigate the
applicability of our results to other materials we perform experiments
with spherical glass beads as well as lead beads, see below.
The cup is mounted on a B\&K shaker and subjected to a vertical
sinusoidal vibration at frequency $\omega=2 \pi f$, with the frequency
scan from sub kHz to 10 kHz. Since the bulk behavior depends on the
contacts between the grains, the way of filling the granular medium in
the container is important. Prior to the measurement, we pour the
grains into the cup and then shake the cup near its resonance
frequency (see below) at large amplitudes up to 3g for 30 minutes to
allow the compaction of the material.  During the measurement, the
granular medium is driven with cup acceleration of approximately 2 to
3 $ms^{-2}$. The amplitude of vibration is approximately 1 $\mu$m for
the lower frequencies and decreases to below 1 nm for the higher
frequencies.

Force $F(\omega)$ and acceleration $a(\omega)$ are independently
measured with a force gauge and an accelerometer attached to the
bottom of the cup. The signals are analogically filtered with a
lock-in amplifier, which is synchronized with the driving signal.
These filtered signals are then digitally recorded.  The force
gauge is sandwiched between the shaker and the cup container. At
each frequency, the force and acceleration are averaged over a 500
ms or longer time window. Prior to this time window, there is a 3
sec waiting time following the onset of the driving frequency in
order to allow the system to reach a steady state, and achieve
accurate measurements. Since $F(\omega)$ and $a(\omega)$ are the
net force exerted on and the response of the cup-granules
assembly, their ratio, the effective mass, is independent of the
shakers frequency response. Taking into account the mass of the
empty cup, $M_c$, we have $ \tilde{M}(\omega) + M_c =
\frac{F(\omega)}{a(\omega)},$ where, $\tilde{M}(\omega) \equiv
M^{'}(\omega)+iM^{''}(\omega)$, reflects the partially in-phase,
$M^{'}$, partially out-of-phase, $M^{''}$, motion of the
individual grains, relative to the driving force.  Thus $M^{'}$
carries information about the inertia and elastic response while
$M^{''}$ relates to attenuation of particle response.

\begin{figure}[b]
\centerline{
\resizebox{8.5cm}{!}
    {\includegraphics{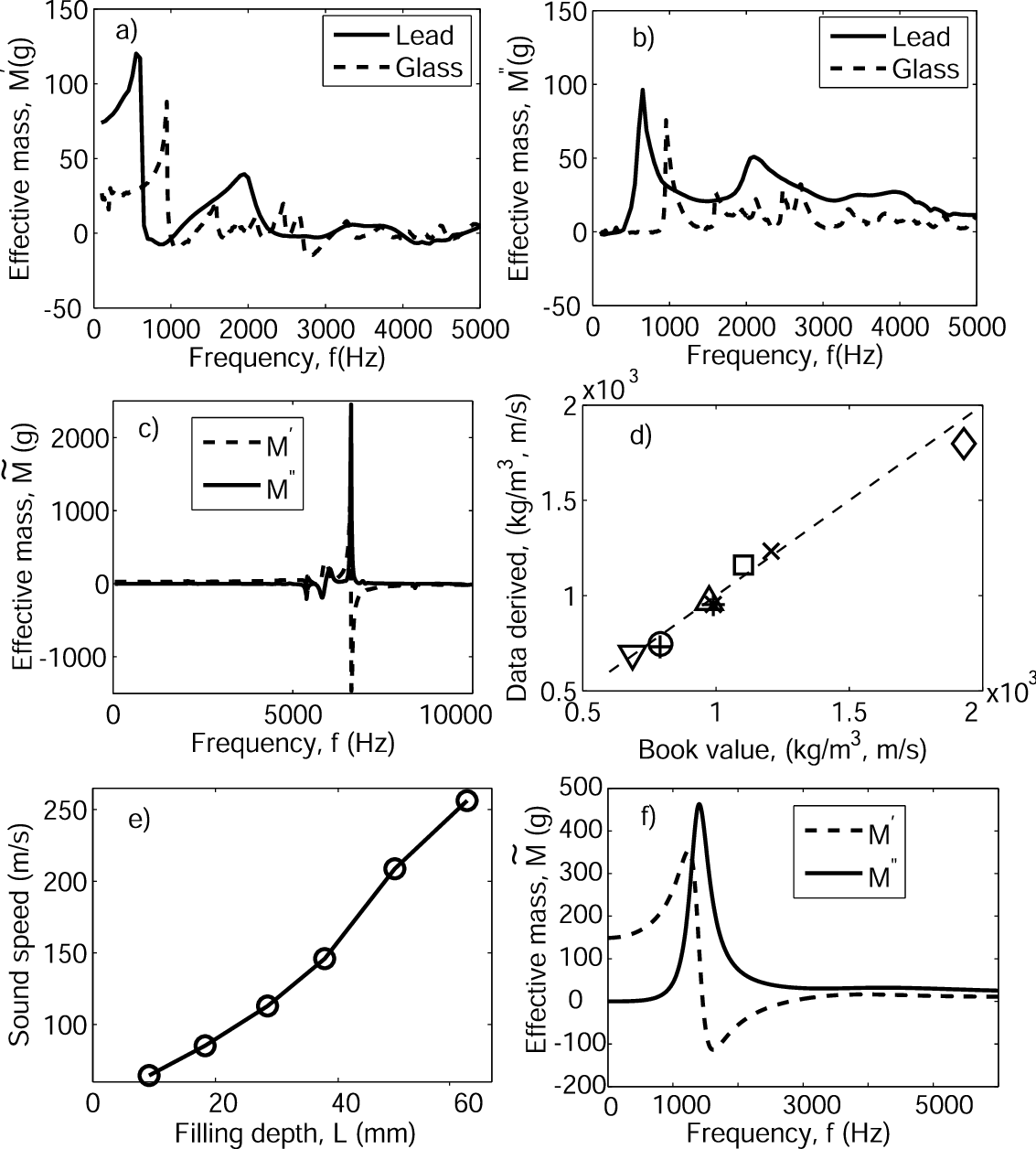}}
}
 \caption{(a) Real ($M^{'}$) and (b) Imaginary part ($M^{''}$) of effective
  mass for spherical lead
and
  glass
  beads.  (c) Effective mass of fluorocarbon fluid. (d) Crossplot of
  measured values and book values of densities and sound speeds for
  various liquids: Methanol (o and $\square$, first symbol density,
  second sound speed), Ethanol (+ and $\times$), Fluorocarbon fluid
  ($\diamond$ and $\triangledown$), Silicon
  oil, (viscosity 10 mPa-s, $\vartriangle$ and $\ast$). (e) Effective
  speed of sound in tungsten granules as a function of filling depth,
  $L$, for constant cavity diameter, $D=25.4$mm.
(f) Continuum model prediction of $
  \tilde{M}(\omega)$.}
\label{fluid_grains}
\end{figure}

Fig. \ref{experiments} shows the result for a frequency sweep of
$\tilde M(\omega)$. We note that in the low frequency limit,
$M^{'}(\omega\to 0)\to 149$g, the static mass of the grains.  More
interestingly, there is a relatively sharp resonance peak around
$f_0\equiv \omega_0/2\pi = $1.8 kHz and a relatively broad tail
that diminishes with increasing frequency. $M^{''}$ exhibits a
broad peak centered at $w_0$. We have conducted the experiments
with other granular materials. Fig. \ref{fluid_grains}(a)-(b)
shows the results for spherical lead beads of the size 180-250
$\mu$m and glass beads of 150 $\mu$m. These results are
qualitatively similar to those of tungsten particles, in both the
real and imaginary parts; the tungsten particles exhibit larger
effects.

We also applied the effective mass technique to liquid samples.
Fig. \ref{fluid_grains}c shows a typical plot for the effective
mass of one such liquid (fluorocarbon fluid, density 1800
kg/m$^3$) as measured by the effective mass technique.
$M^{'}(\omega)$ shows a sharp resonance while the $M^{''}(\omega)$
curve is almost zero over the whole frequency range (except near
resonance) indicating negligible dissipation.  We can extract a
value for the effective compressional sound speed from the
resonance frequency as $v_s(L) = 4 L f_0$, where L is the depth of
the medium. The values of density and sound speed for a variety of
liquids, as measured with the effective-mass technique, are
cross-plotted against those determined by more conventional means
\cite{crc} in Fig. \ref{fluid_grains}d. The density of the liquids
is measured using Archimedes principle and buoyancy (accuracy of
0.1\%). The sound speed is measured with ultrasonic (centered at
500 kHz) through-transmission method (accuracy of 1\%).  The grain
density of the tungsten sample is measured with a Micromeritics
helium pycnometer (accuracy 1\%) and the bulk density is
calculated from the measured weight of tungsten grains filling a
cup of a known volume. Volume fraction is then estimated from the
grain and bulk density. The results in Fig. \ref{fluid_grains}d
suggests that our technique for measuring $\tilde M(\omega)$ is an
accurate one.


We investigate how the effective sound speed changes with the
filling depth $L$, based on the peak frequency $f_0$.  The sound
speed of the granular media is estimated based on the hypothesis
of 1/4 wavelength resonance frequency. Fig. \ref{fluid_grains}e
shows the trend of greater speed with greater depth.  The values
we are reporting are of the same order of magnitude as those
reported in the literature \cite{cremer,bourinet2,liu,shield}.
In the range studied here, the linear relation $v_s(L) \sim L$
seems to hold up to $L=61$ mm as seen in Fig. \ref{fluid_grains}e.

We may model the general features of Fig. \ref{experiments} in terms
of a simplified continuum model. The granular medium is considered to
be a lossy fluid, with negligible viscous effects at the walls.  The
acoustic pressure in the fluid is described by a one-dimensional
model: $P(z)= A\sin(qz) + B\cos(qz)$. The boundary conditions are: (a)
The displacement of the fluid at $z=0$ matches that of the cup,
$u(z=0) = u_{\rm cup}$. (b) The acoustic pressure at the top of the
fluid vanishes, $P(z=L) = 0$.  Thus, $P(z) = u_{\rm cup} \rho \omega^2
[\sin(qz) - \tan(qL) \cos(qz) ]/q$. The force the cup exerts on the
fluid is $F=\pi a^2 P(z=0)$ and the acceleration of the cup is
$-\omega^2 u_{\rm cup}$ from which we obtain the effective mass:
\begin{equation}
\tilde{M}(\omega) = \pi a^2  \sqrt{\rho K}/\omega \tan(qL),
\label{M}
\end{equation}
where $K= K_0[1-i \omega \xi]$ is the (lossy) bulk modulus of the
medium, and $q = \omega \sqrt{\rho/K}$ is the complex wave vector with
$\rho$ the density of the medium. This model
shows that resonance peaks, as seen in Fig. \ref{fluid_grains}f,
occur when $L$ equals odd multiples of 1/4 wavelength of the
acoustic wave in the medium, with the main resonance peak at
$f_0=1/(4 L) \sqrt{K_0/\rho}$.  Encouraged by these results, we
interpret the main resonance in Fig.  \ref{experiments} as being a
1/4 wavelength resonance of the compressional sound speed.



Next, we use the features in $\tilde M(\omega)$ to investigate the
dissipation in the granular medium. The dissipation mechanisms
can be grouped into two categories: 1) Local damping throughout the
contact network that includes asperity deformation and bulk
attenuation \cite{poschel2}, and wetting dynamics in the liquid
bridges caused by relative motion of neighboring particles
\cite{1}. Wetting dynamics also results in greater contact stiffness
\cite{damour}.  If large interparticle tangential displacements occur
then, there exist dissipation through Coulomb friction as well. 2)
Global damping due to drag caused by particle motion in a surrounding
viscous medium \cite{wolf-md}.

%


Next, we perform further experimentation by changing the ambient
conditions.  Ambient humidity can modify the dissipation and the
stiffness at the grain to grain surface contacts \cite{damour}.
For instance, small amounts of liquid greatly affects the maximum
angle of stability in sandpiles \cite{barabasi}, while capillary
action of the interstitial liquid results in increased cohesion
\cite{damour}, modified friction \cite{riedo} and aging in the
angle of repose of granular materials \cite{bocquet}.  We place
the effective mass apparatus in a chamber where humidity is
controlled by first using an aqueous saturated salt solution
\cite{restango} and subsequently a dessicant. We measure the
relative humidity as the ratio of the partial pressure of water to
the saturating vapor pressure $H_{\rm
  R}=p_{\rm vap}/p_{\rm sat}$.

\begin{figure}[t]
\centerline { \hbox{ \resizebox{9.5cm}{!} {
\includegraphics{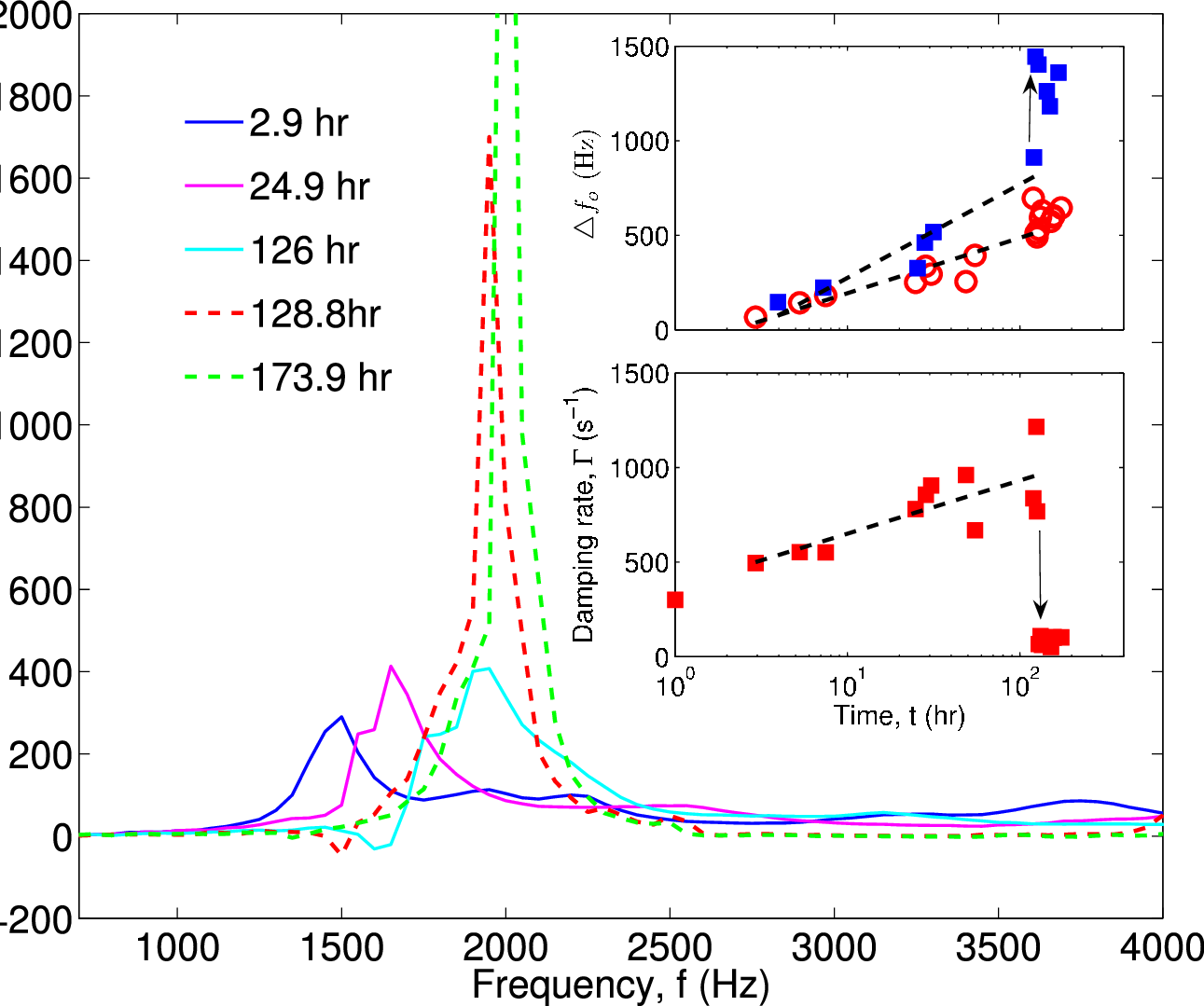}}}}
\caption{Imaginary part of $\tilde M(f)$ measured under humidity
  controlled conditions (the real part shows similar behavior). The
  solid lines represent measurements under $H_{\rm R}=75\%$ and the
  dashed lines are for the dry state, $H_{\rm R}=0\%$. Top inset:
  Frequency shift vs time for $H_{\rm R}$=75\% (o) and 95\%
  ($\square$). Bottom inset: Damping rate ($\Gamma$) vs time for
  $H_{\rm R}$=75\%. The dashed lines in the insets are logarithmic
  fits. The up and down arrows indicates the large variations in
  stiffness and damping upon drying, respectively.}
\label{humidity}
\end{figure}

We follow a humid-dry cycling process as follows: we start from
room humidity $H_{\rm R}= 26\%$ at $t<0$, and then we humidify at
$t=0$ to $H_{\rm R}= 75\%$ until $t=126$ hr, and then we dry to
$H_{\rm R} = 0\%$ until $t=174$ hr. Fig. \ref{humidity} shows the
effective mass measurements under the humidity controlled
environment carried out at various time intervals. We observe a
very slow shift in resonant frequency during the humidification
cycle, followed by a further positive shift upon drying. The
effect is more important when we repeat the same cycling but to
$95\%$ humidity.

During humidification, we observe an aging phenomenon as evidenced by
the approximate logarithmic behavior of the shift in resonant
frequency $\Delta f_0(t)$ (top inset of Fig. \ref{humidity}), $\Delta
f_0(t) = C(H_{\rm R}) \log(t/t_i),$
where $ C(H_{\rm R})$ is a constant that increases with humidity,
$ C(75\%)=292$ Hz and $C(95\%)=494$ Hz and $t_i\approx 2$ hr for
both measures.

Furthermore, the humid system shows a large increase in the
attenuation as illustrated by the increase in damping rate
$\Gamma$ (bottom inset of Fig. \ref{humidity}). Based on the
experimental effective mass data,
we search for poles in $\tilde{M}(\tilde\omega)$ for $\tilde\omega$ in
the complex plane. Once the complex resonance frequency
($\tilde{\omega}_{0}=\omega_{0}^{'}+i\omega_{0}^{''}$) is found, the
damping rate is given by $\Gamma=-\omega_{0}^{''}$.
The damping rate at ambient humidity is 300 $s^{-1}$, which rises to
1200 $s^{-1}$ under prolonged humidification
(see bottom inset of Fig. \ref{humidity}). The damping data suggests
logarithmic increase with time.

A possible interpretation of the observed behavior suggests that
contact forces through capillary liquid bridges are important for
damping, stiffness and aging. As demonstrated in \cite{bocquet},
the humidification process leads to the formation of capillary
bridges around the roughness asperities at the interparticle
contact surfaces. The aging effect observed in the logarithmic
variation of the frequency shift with time can be interpreted as
originating from capillary condensation. The key quantity to
calculate is the time dependence of the number of wetted
asperities $n_a(t)$.  Capillary condensation at the surface
asperities can be seen as a thermally activated process following
a model proposed in \cite{bocquet}.  The energy barrier to form a
liquid bridge of cross sectional area $A$ between two asperities
separated by a gap $h$ is $\Delta E = \Delta \mu \rho_{l} Ah$,
where $\rho_{l}$ is the molecular density of the condensing
liquid, $\Delta \mu = k_{\rm B} T \ln(1/H_{\rm R})$ is the
undersaturation of the chemical potential, and $T$ the room
temperature. The activation process follows an Arrhenius dynamics
such that the time to create a bridge is $\tau = t_0 \exp(\Delta
E/ k_{\rm B} T)$, with $t_0$ the microscopic condensation time.

Due to the roughness of the particles we expect a distribution of gaps
$P(h)$, which leads to a distribution of energy barriers, $P(\Delta
E)$. The calculation of the probability distribution $P(\tau)$ is then
analogous to obtaining the probability of the time to escape from a
trap with energy barrier $\Delta E$ for a thermally activated particle
in a random energy potential. Such a process corresponds to the ``trap
model'' used to describe the phenomenology of glasses
\cite{bouchaud}. The probability $P(\tau)$ satisfies $P(\tau) d\tau=
P(h) dh$.  While the probability $P(h)$ is unknown, we expect it to be
a bounded distribution determined by the surface roughness of the
particles. The two most common cases lead to the same result: for an
exponential or Gaussian distribution of gaps, we find $P(\tau) \sim
\tau^{-1}$.

The number of activated wetted asperities at time $t$ is $n_a(t)
\propto \int_{t_0}^{t} P(\tau) d\tau$.  Then $n_a(t) \approx
\log (t/t_0)$.
Liquid bridges induce a significant cohesion between the particles
which for loose grains under gravity and large humidity outweigh the
weight of the grains. Therefore, the logarithmic increase in $n_a$
leads to an additional stiffness with the concomitant increase in the
compressional sound speed. Since the resonance frequency is
proportional to the stiffness of the contact,
the logarithmic aging can be interpreted as arising from the wetted
asperities.

Upon drying the particles to $H_{\rm R} = 0\%$ after humidification,
the physics is very different from the humid process.  As shown by
the experiments of \cite{damour}, the capillary rings dry out and
the asperities are sucked down by the increasing capillary
pressure in the evaporating film leading to an even larger
increase in stiffness of the contacts. Thus, in the dried out
state there is much more solid on solid contact and the sound
speed is increased.  This increase in stiffness upon drying
results in even larger positive shifts in resonant frequency as
seen in the top inset of Fig. \ref{humidity}.

We also find important changes in the dissipation in the dry state.
Upon drying, the decrease in overall attenuation is evident from
the much sharper resonance in $M^{''}$ (we find a very large mass
at resonance $M^{''}(f_0)\approx 3500$ g). Accordingly, the
damping rate decreases to $\Gamma=50$ $s^{-1}$ in the dry case
from $\Gamma=1200$ $s^{-1}$ at $H_{\rm R}=75\%$ (see bottom inset
of Fig. \ref{humidity}). In the driest state, no liquids bridges
are left and the dissipation is purely through asperities
deformation with the concomitant lower damping strength. These
results suggest
that, for the granular media studied here
when humid conditions prevail (including ambient), liquid bridges
provide larger attenuation of acoustic waves than the viscoelastic
dissipation in the bulk of the particles or viscous global damping.

In conclusion, we have demonstrated the usefulness of the effective
mass technique to study the dissipation of acoustic waves, stiffness
and aging in granular media. Experiments and theoretical models
suggest that capillary condensation at the interparticle contacts is
an important mechanism for dissipation and aging.
The kinetics of aging can be modeled in terms of a trap model for the
formation of liquid bridges at the asperities.  The large variations
in dissipation found in this study demonstrate the conditions for
effective particle damping and are relevant to a variety of
applications for optimizing attenuation of structure-borne acoustic
waves.

{\it Acknowledgments. }
We are grateful to L.  McGowan for technical assistance.  We
acknowledge financial support from DOE, Geosciences Division.

\end{document}